# Social Media Impact on Website Ranking

**Dushyant Vaghela**[1]
http://explorequotes.com
Department of Computer Engineering
RK University, Rajkot, India.

*Abstract*—Internet is fast becoming critically important to commerce, industry and individuals. Search Engine (SE) is the most vital component for communication network and also used for discover information for users or people. Search engine optimization (SEO) is the process that is mostly used to increasing traffic from 'free', 'organic' or 'natural' listings on search engines and also helps to increase website ranking. It includes techniques like link building, directory submission, classified submission etc. but SMO, on the other hand, is the process of promoting your website on social media platforms. It includes techniques like RSS feeds, social news and bookmarking sites, video and blogging sites, as well as social networking sites, such as Facebook, Twitter, Google+, Tumblr, Pinterest, Instagram etc. Social media optimization is becoming increasingly important for search engine optimization, as search engines are increasingly utilizing the recommendations of users of social networks to rank pages in the search engine result pages. Since it is more difficult to tip the influence the search engines in this way. Social Media Optimization (SMO) may also use to generate traffic on a website, promote your business at the center of social marketing place and increase ranking.

## I. INTRODUCTION

Social activities and thinking outline have usually been spread via radio, newspapers and television. However, a new trend has evolved: social media. Social media is the approach to socialize. In early, the web was plays a vital role to retrieval of discrete information but now web has become a social place that connect people worldwide. Now use of the social media becomes common most of users and business person as different ways. [1]

Social media is two-way messaging between company and follower/ People. The term 'social media' can be dividing into two parts. Media usually refers to publicity and the communication of information through publications and channels. Social involve the interaction of user within a group or community. [2]

Any website which permit user to like, dislike, tweets and share their content, and allow making community can be classified as a social media. Some social media platforms are: (1) Social Bookmarking site such as delicious, stumble upon. (2) Social Networking sites such as Facebook, Google+. (3) Photos and video sharing such as Flickr and YouTube (4) Microblogging site such as Twitter and (5) Blogs such as WordPress and Blogger.
This is a good Wiki explanation of that direction…

Social media optimization is becoming increasingly important for search engine optimization, as search engines are increasingly utilizing the recommendations of users of social networks such as Facebook, Twitter, and Google+ to rank pages in the search engine result pages... since it is more difficult to tilt the scales search engines in this way, search engines are putting more stock into social search. This, coupled with increasingly personalized search based on interests and location, has significantly increased the importance of a social media presence in search engine optimization. [3]

Rohit Bhargava was first used the term "Social Media Optimization" in 2006 and also discovers five rules. (1) Increase your link ability (2) Make tagging and bookmarking easy (3) Reward inbound links (5) help your content travel (6) Encourage the mash up [4]

## II. SOCIAL MEDIA PLATFORM

### A. Type of Social Media and its Factors

It's one thing to know that social factors are increasingly affecting search engine rankings, but it's another thing entirely to see what integrations are already in place and influencing results pages. There are some social media Platform describes below.

*1) Facebook*

Facebook is the largest social platform of the social networks, with worldwide usage approaching 700 million users as of June 2011. This represents nearly 10% of the total world population. Facebook has the largest audience of any social network and is certainly worth including as a key part of your social media strategy. People can implement links on their Facebook pages. If the user's news page is public, these links will be visible to the search engines and can and will be counted as links. Not only that, when people share links they will also show up on the news pages of some of their friends. Publishers can encourage the sharing of their content by placing a Facebook Share button on their pages. [5]

In the current Google Social Search implementation, content shared by Facebook contacts receives preferential positions in the search engine results, as well as with a picture of the person who shared it. By increasing the social proof of these results, Google increases the chances that this content will receive clicks.

Facebook is the best social media site to *increase traffic to your website* by using the quality. Facebook groups, building up your Facebook page likes, and creating engagement that you create on your wall and the Facebook page wall. Images post work really well. So add images to your profile and the Fan Page wall every day. Add targeted people as friends and always message them with a message that positively encourages them to like your Facebook page. [7]

*2) Twitter*

One of the clearest integrations between social media and the search engines is the Twitter.

In April 2011, Twitter was in use by approximately 7% of the US population and this number was expected to grow to 11% by the end of 2011. Twitter market share may be small compared to the penetration of Facebook, but it certainly includes a large number of influencers that can help create visibility for your organization.

One of the unique ways to clutch traffic from Twitter, and it was by following new people and getting them to visit your blog via your profile links. Engage with as many other tweeters and you may have them become a visitor to your blog! [7]

*3) Google +*

Google+ was first announced on June 28, 2011. In Google's earnings announcement on January 12, 2012, it indicated that the social network had 90 million users, and that 60% of those people were active on the site every day. This rapid growth and broad adoption suggests that Google itself may be a strong player in the social field of the future and it now has a major impact on Google search rankings.

Google+ allows sharing in a manner similar to what you can do on Facebook. Once you insert a link into a post on the site, Google+ recognizes the URL and extracts content.

If you want to getting traffic from Google +, you should provide the name and contribute to Google +'s Communities. Google +'s communities superb way to more traffic to your site. Make sure you don't post into them without some sort of contribution or engagement you do for them, no matter how small. Make a practice to use the right Hashtags (#) when you are posting in the communities. [7]

*4) LinkedIn*

As of January 2011, LinkedIn had more than 101 million members worldwide. This makes it a powerful social network. LinkedIn can be used to build a strong network of connections, and paid versions of the service allow its. In Mail functionality to be used to send unsolicited emails to members; if done judiciously, this can be used to initiate new relationships with influencers of interest. It can also use to make a strong profile of users. [5]

*5) YouTube*

YouTube can be thought of as a search engine, but it is also a social network. People love to share videos, and this is happening in volume. Video hosting service from YouTube is a social networking website, which almost every person or organization with Internet access can upload videos that can be seen by the general public almost immediately. As the world's largest video platform, YouTube has had an impact in many areas, with some individual YouTube videos directly happens worldwide have formed. [5]

Below there are some factors used by several web services and tools that affect the authority, the trust of social media links and also assess Social Media reputation [11]

| Sr. No | Factors |
|---|---|
| 1 | Followers on Twitter |
| 2 | Fans on Facebook Fan Pages |
| 3 | Shares & Likes on Facebook |
| 4 | The number of tweets & RTs on Twitter (Reply Tweets of your Tweet) |
| 5 | ratio between Followers-Friends |
| 6 | The authority of the people that follow you |
| 7 | The authority of the people that share your content |
| 8 | The average quality of your previously shared messages |
| 9 | The number of unique mentions/shares (similar to Link Diversity) |
| 10 | The rate and the source of links that you share |

Table. 1: Social media Factors Moreover search engines are expected to use Social Media buzz to determine:

- How a particular post is fresh
- Is Particular page still useful/valid/up-to-date
- What are the current trends
- For the particular post which terms are related
- They are likely to use the keywords and the hash tags that can be found in the content of the shared messages.[11]

*B. Why Rely On Social Media?*

If you are still wondering why Google is pushing so hard with its new product Buzz, it is because it wants in on social traffic. For many sites on the Web, social traffic coming through Facebook, Twitter, and Myspace is beginning to rival, and in some cases overtake, search traffic as the single biggest source of traffic. This traffic comes from shared links, photos, and videos. By its own numbers, 5 billion pieces of content are shared on Facebook every week.

Social media used in business, ranking and generate traffic.

The top two benefits of social media marketing are increasing exposure and increasing traffic. A significant 89% of all marketers indicated that their social media efforts have generated more exposure for their businesses. Increasing traffic was the second major benefit, with 75% reporting positive results.

You can use social media to drive traffic to your website. Many companies from small to large are using social media to drive traffic to their websites and products. As a part of study made 71% percent of social media users are more likely to purchase from a brand they follow online. Social media is increasing every day and companies not using social media will be left behind. Finding different ways to promote your website through social media is very crucial to the success of your site. [6]

III. IMPACT OF SOCIAL MEDIA There are many Survey by different SEO Company / expert about social media such as Facebook, Twitter and Google +.

Fig. 1 represents Summary of Search metrics SEO Ranking Factors – Rank Correlation 2013. In the Figure 1, the correlation coefficient is plan on the x-axis, which measures its size. The longer/shorter the bar in the positive part of the x-axis that represent the higher/lower the correlation between that particular factor and a good Google ranking. Factors with a correlation factor of zero do not have any considerable correlation with Google's results. [8]

In this chart, there are some factor of social networking such as Facebook, Twitter and Google + with its ranking possibility which platform is more important.

It is noticeable that social signals correlate well with better rankings and content quality as well as the number and diversity of backlinks seem to have a huge impact on search result rankings. [10]

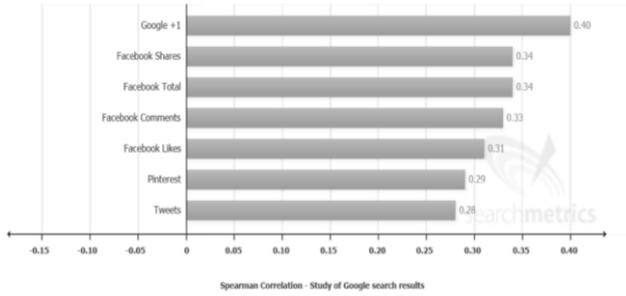

Fig. 1: Study of Google Search Result

Below you will also find the hottest questions that have been answered from interview along with the replies of each Search Engine: Danny Sullivan asks to both Google and Bing during interview whether the data of Twitter and Facebook affect the regular search Result or not. [9]

| Sr.No | Question | Google | Bing |
|---|---|---|---|
| 1 | Are Regular Search Results affected by social media buzz? | Yes. It is used as a signal especially for news. | Yes it is a signal. Some weight is passed and regular results are affected. |
| 2 | Are Social/Realtime Search Results affected by buzz? | Heavily affected | Heavily affected, Authority metrics are used to determine the hot posts. |
| 3 | Are Twitter links taken into account (aka do they pass link juice)? | In some limited situations the data are used. | The data are used. The weight depends on how often a link is posted, the number of tweets & retweets and the authority of the people that post it. |
| 4 | Are Facebook links taken into account? | The shared links from Fan pages are treated similarly to Twitter links. No links from personal walls are used. | Only the publicly shared links from Fan pages and Profiles are taken into account. |
| 5 | Is there an Authority Rank for Twitter Profiles? | Yes, the author quality is measured. Google calls this Author Authority. | Yes. Several metrics are taken into account. Bing calls this Social Authority. |
| 6 | Is there an Authority Rank for Facebook Profiles? | They are treated similarly to Twitter Profiles. Note: Normally if this is true, they measure only public profiles like Fan pages. | No. They take into account only the Facebook data that are publicly available. |
| 7 | Does the Twitter Authority have an impact on the weight of the links that are shared? | Yes. The weight depends on the authority of the person. The Author Authority is used in limited situations and it is independent of PageRank. | The Author Authority affects the link weight. |
| 8 | Does the Facebook Authority have an impact on the weight of the links that are shared? | Similarly to Twitter. The weight of each link depends on the authority of the profile. | They don't use Facebook authority. Instead in order to find if something is really popular they compare Twitter & Facebook shares. |

Table. 2: Google and Bing Interview

Based on the interview by Danny Sullivan, Chief Editor of Search engine land which confirms the doubts of many SEOs: Social Media does affect the organic rankings of Google and Bing. Some social media play a vital role for search engine but does it directly influence the ranking. IV. CONCLUSION

In this review paper, I concluded that Social media and user engagement have become significant factors for those practicing SEO. Links may still be the strongest source of signals to search engines. A balanced SEO strategy needs to take social media, user data, and document analysis techniques into account. Publishers must consider these four areas as a part of their web promotional strategies: (1) Participating in social media communities (2) Providing an engaging user experience (3) Offering unique and differentiated content (4) Building a brand.

I focused on social media platform such as Social Networking site, Social Bookmarking, microblogging, all have different functionality and also the use of social media goal to improve business, increase / generate traffic to your website and ranking in SERP. Some social networking site factor most likely used such as Facebook Share, Twitter tweets but some are less likely used such as Facebook Like, Google+ for increase traffic and ranking.

Based on the survey, Impact of Google + is very strong compare to other to ranking but Facebook like, share and twitter tweet are helps to generate traffic but continuously engagement is required. If we focus on Facebook's like, comment and Google+, tweets and social bookmarking factor, we can easily promote business, generate traffic and improve ranking in SERP.